\documentclass{article}
\usepackage{spconf,amsmath,graphicx,hyperref}

\usepackage{nth}
\usepackage{bbm}
\usepackage{graphicx}
\usepackage{color}
\usepackage{cite}
\usepackage{amsmath}
\usepackage[caption=false]{subfig}
\usepackage{amssymb}
\usepackage{comment}
\usepackage{mathtools}
\usepackage{tabulary}
\usepackage{acronym}
\usepackage{upgreek}
\usepackage{algorithm}
\usepackage{algpseudocode}
\usepackage{placeins,flushend}

\title{Integrating Stacked Intelligent Metasurfaces and Power Control \\for Dynamic Edge Inference via Over-The-Air Neural Networks}
\name{Kyriakos Stylianopoulos and George C. Alexandropoulos
\thanks{This work has been supported by the SNS JU project 6G-DISAC under EU's Horizon Europe research and innovation programme under Grant Agreement number 101139130. AWS resources were provided by the National Infrastructures for Research and Technology GRNET and funded by the EU Recovery and Resiliency Facility.}
}

\address{Dpt. of Informatics and Telecommunications,
National and Kapodistrian University of Athens, Greece\\
emails: \{kstylianop, alexandg\}@di.uoa.gr}
\begin{document}
%
\maketitle
\begin{abstract}
This paper introduces a novel framework for Edge Inference (EI) that bypasses the conventional practice of treating the wireless channel as noise. We utilize Stacked Intelligent Metasurfaces (SIMs) to control wireless propagation, enabling the channel itself to perform over-the-air computation. This eliminates the need for symbol estimation at the receiver, significantly reducing computational and communication overhead. Our approach models the transmitter-channel-receiver system as an end-to-end Deep Neural Network (DNN) where the response of the SIM elements are trainable parameters. To address channel variability, we incorporate a dedicated DNN module responsible for dynamically adjusting transmission power leveraging user location information. Our performance evaluations showcase that the proposed metasurfaces-integrated DNN framework with deep SIM architectures are capable of balancing classification accuracy and power consumption under diverse scenarios, offering significant energy efficiency improvements.  
\end{abstract}
\begin{keywords}
Goal-oriented communications, edge inference, stacked intelligent metasurfaces, over-the-air computing, deep learning.
\end{keywords}

\section{Introduction}
The deployment of reflecting or diffractive metamaterials~\cite{Metasurfaces_Review_2016, RIS_Survey_George} into the propagation environment enables the dynamic manipulation of wireless channels.
By applying phase and, possibly, amplitude control on traveling waves, such programmable MetaSurfaces (MSs) can serve as enablers in a plethora of communication and sensing applications~\cite{9693982,Counteracting,GAHenkHui,6G-DISAC-magazine}.
The role of MSs has been also recently extended to Over-the-Air (OTA) computational tasks~\cite{RICS_2025, OTA25_Rahimian}, building on the framework of optical computing~\cite{Metamaterials_Operations_Science_2014}. Interestingly, such principles have been exploited in controlled, free-space, environments to implement OTA Deep Neural Networks (DNNs)~\cite{XYN18_Diffractive_DNN, LML23_D2NN}, mainly using Stacked Intelligent Metasurfaces (SIMs)~\cite{AXN23_SIM}.

The union of MS-based OTA and Deep Learning (DL) has great potential in the domain of Goal-Oriented Communications (GOC) for Edge Inference (EI).
Under GOC, a DNN typically performs Joint Source-Channel Coding (JSCC)~\cite{DeepJSCC, DeepSC} so that its layers are split across the transceivers and the channel is treated as a hidden, yet uncontrollable, OTA-realized DNN layer.
EI then evaluates a data-driven objective (e.g., instance classification) based on the decoded data, which serves as an objective function for the End-to-End (E2E) system.
SIM components have been deployed at the transceivers in~\cite{GJZ24_SIM_TOC} to endow the GOC devices with analog DL processing, while a nonlinear eXtremely Large (XL) MS at the Receiver (RX) side was used in~\cite{Stylianopoulos_MIMO_ELM} to capitalize on rich scattering and implement a single-layer feedforward network with proven universal approximation capabilities.
Additionally, the Transmitter (TX)-MS-RX system was optimized in~\cite{Gunduz_Layer_Approximation} to approximate a hidden DNN layer under an OTA GOC scenario.
In the very recent work~\cite{Stylianopoulos_GO}, a comprehensive study of MS-Integrated Neural Networks (MINNs) was conducted, where the MS-parametrized channel played the role of a cascade of hidden network layers, either directly trainable through backpropagation or controllable via DNN modules that observed the current channel states.

\begin{figure}[t]
    \centering
    \includegraphics[width=0.9\linewidth]{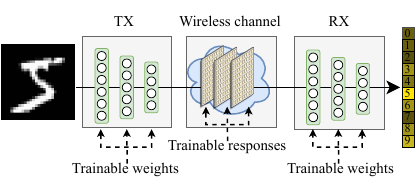}
    \vspace{-0.3cm}
    \caption{MINN architecture with tunable MS-based layers.}
    \label{fig:arch}
\end{figure}

Despite notable results in OTA computational offloading offered by the previous works~\cite{XYN18_Diffractive_DNN,LML23_D2NN,GJZ24_SIM_TOC,Stylianopoulos_MIMO_ELM,Gunduz_Layer_Approximation,Stylianopoulos_GO}, key attributes of MINNs regarding performance limits, systemic benefits, and resource utilization call for detailed investigation.
In particular, the roles of the number and size (in terms of unit elements) of MSs required to approximate digital DNNs, as well as the necessary power budgets and Multiple-Input Multiple-Output (MIMO) antenna configurations sufficient for successful EI, especially under dynamic fading, are still unexplored. In this paper, we extend the MINN framework~\cite{Stylianopoulos_GO} of treating the wireless channels as a controllable medium for DL computations (as illustrated in~\autoref{fig:arch}) towards the investigation of the latter. We consider exclusively SIM layers whose element responses are trainable, equivalent to DNN weights.
Upon training, the responses remain fixed, despite changes in channel states, allowing for important hardware simplifications due to the absence of dynamic control circuits on the MSs.
Furthermore, the MINNs are integrated with a DNN module that is tasked to control the transmission power based on the user's position. The presented performance evaluation showcase trade-offs between antenna and MS elements, as well as energy consumption and classification accuracy.

\vspace{-0.2cm}
\section{System Model and Design Objective}
\vspace{-0.2cm}

\subsection{System Model}
We consider narrowband downlink MIMO transmissions where a SIM of $L$ planar layers of $N$ elements each (i.e., $N_{\rm sim} \triangleq LN$ total elements) is placed at the vicinity of a static TX with $N_{\rm t}$ antenna elements, while an $N_{\rm r}$-element mobile RX traverses an area of static clutter, represented by $K$ individual scatterers.
Following the Saleh-Valenzuela model~\cite{SV_Model}, the TX-RX, $\mathbf{H}_{t,{\rm r}} \in \mathbb{C}^{N_{\rm r} \times N_{\rm r}t}$, and SIM-RX, $\mathbf{H}_{\rm{m'},{\rm r}r} \in \mathbb{C}^{N_{\rm r} \times N_{\rm sim}}$, channels can be expressed as follows:
\begin{align}
  \mathbf{H}_{{\rm t},{\rm r}} &\triangleq \sqrt{\frac{N_t N_{\rm r}}{K}} \sum_{k=1}^{K} P^{\rm L}_{{\rm t},k} P^{\rm L}_{k,{\rm r}} \alpha_k \mathbf{a}_{\rm r}(\theta_{{\rm r},k}) \mathbf{a}^H_{\rm t}(\theta_{{\rm t},k}), \label{eq:channel-tx-rx}\\
  \mathbf{H}_{{\rm m'},{\rm r}} \! &\triangleq \! \sqrt{\frac{N N_{\rm r}}{K}} \! \sum_{k=1}^{K} \! P^{\rm L}_{{\rm m'},k} P^{\rm L}_{k,{\rm r}} \alpha_k \mathbf{a}_{\rm r}(\theta_{{\rm r},k}) \mathbf{a}^H_{{\rm m'}}(\theta_{{\rm m'},k}), \label{eq:channel-ris-rx}
\end{align}
while the TX-SIM channel is modeled as a Line-of-Sight (LoS) channel due to the close proximity of the nodes:
\begin{equation}
      \mathbf{H}_{{\rm t},{\rm m}} \triangleq \sqrt{N_{\rm t} N} P^{\rm L}_{{\rm t},{\rm m}}  \mathbf{a}_{\rm m}(\theta_{{\rm t},{\rm m}}) \mathbf{a}^H_{{\rm t}}(\theta_{{\rm t}, {\rm m}}) \in \mathbb{C}^{N \times N_{\rm t}}.
\end{equation}
In the above, ${\rm t}$, ${\rm r}$, ${\rm m}$, ${\rm m'}$ signify the TX, RX, first and last SIM layers, respectively.
Moreover, $P^{\rm L}_{i,j}$ denotes the free-space pathloss between the $i$ and $j$ nodes or scatterers, $a_k$ is the complex path gain when reflected by the $k$-th scatterer, $\mathbf{a}_i(\cdot)$ is the array response vector of the $i$-th node, while $\theta_{i,j}$ denotes the angle between the $i$ and $j$ nodes or scatterers.

The element-to-element propagation between consecutive SIM layers is governed by geometric optics due to their dense placement~\cite{AXN23_SIM, GJZ24_SIM_TOC, Stylianopoulos_GO}.
Given elements $n$ and $n'$ ($1 \! \leq \! n,n' \! \leq \! N$) with distance $d_{n,n'}$ and area $S_M$ from layers $l$  and $l\!-\!1$ ($2 \! \leq \! l \! \leq \! L$) of distance $d_M$, the propagation matrix $\mathbf{\Psi}_l\! \in \! \mathbb{C}^{N \times N}$ is:
\begin{align}
    [\mathbf{\Psi}_l]_{n,n'} &\triangleq \frac{d_M S_M}{d_{n,n'}^2} 
    \Big( \frac{1}{2\pi d_{n,n'}} - \frac{\jmath}{\lambda} \Big) 
    \exp\left({\jmath 2\pi d_{n,n'}}\right),
\end{align}
where $\lambda$ is the carrier frequency and $\jmath\triangleq\sqrt{-1}$.
The responses of the unit elements of the $l$-th layer $\boldsymbol{\phi}_l$ are modeled as typical idealized unit-amplitude phase shifters, i.e., $[\boldsymbol{\phi}_l]_n \triangleq \exp(-\jmath \vartheta^l_n)$
\footnote{We use notation $[\mathbf{z}]_i$ and $[\mathbf{Z}]_{i,j}$ to denote respectively the $i$-th element of a vector $\mathbf{z}$ and the $(i,j)$-th elements of a matrix $\mathbf{Z}$.}
, where $\vartheta^l_n$ is the controllable phase shift.
By defining $\mathbf{\Phi}_l \triangleq {\rm diag}(\boldsymbol{\phi}_l)$, the overall SIM response can be mathematically expressed via the following matrix:
\begin{equation}
 \mathbf{\Phi}_{{\rm m},{\rm m'}} \triangleq\left(\prod_{l=L}^{2} \mathbf{\Phi}_l \mathbf{\Psi}_l \right) \mathbf{\Phi}_1\in\mathbb{C}^{N \times N}.
\end{equation}
Given a unit-power transmit vector $\mathbf{x} \in \mathbb{C}^{N_t \times 1}$, the transmit power budget $P$, and the Additive White Gaussian Noise (AWGN) vector $\mathbf{\tilde{n}} \sim \mathcal{CN}(\mathbf{0}_{N_r}, \sigma^2 \mathbf{I}_{N_r})$, the baseband equivalent received signal at the RX is finally defined as follows:
\begin{equation}\label{eq:transmit-signal}
    \mathbf{y} \triangleq  (\mathbf{H}_{{\rm m'},{\rm r}} \mathbf{\Phi}_{{\rm m},{\rm m'}} \mathbf{H}_{{\rm t},{\rm m}} + \mathbf{H}_{{\rm t},{\rm r}})\sqrt{P}\mathbf{x} + \mathbf{\tilde{n}} \in \mathbb{C}^{N_{\rm r} \times 1}.
\end{equation}

\vspace{-0.5cm}
\subsection{Problem Definition}
\vspace{-0.1cm}
Consider a training data set of $D$ paired input and inference target observations $\mathcal{D} \triangleq \{ \mathbf{o}^{(i)}, \mathbf{t}^{(i)} \}_{i=1}^D$. In this paper, we treat the E2E XL MIMO system as a mapping function $ \{ \mathbf{\hat{t}}^{(i)}, P^{(i)} \} \triangleq f_{\mathbf{w}}(\mathbf{o}^{(i)}, \boldsymbol{\mathcal{H}}^{(i)})$, parametrized by a weight vector $\mathbf{w}$, that outputs an estimate $\mathbf{\hat{t}}^{(i)}$ for $\mathbf{t}^{(i)}$ as well as the TX power $P^{(i)}$, where $\boldsymbol{\mathcal{H}}^{(i)} \triangleq \{\mathbf{H}_{{\rm t},{\rm r}}^{(i)}, \mathbf{H}_{{\rm t},{\rm m}}^{(i)},\mathbf{H}_{{\rm m'},{\rm r}}^{(i)} \}$ represents the random overall channel state during the $i$-th EI step. 
We propose the following problem formulation for $\mathbf{w}$'s design:
\begin{subequations}\label{eq:problem}
\begin{align}
\min_{\mathbf{w}}\mathcal{L}(\mathbf{w}) &\triangleq \mathbb{E}_{\boldsymbol{\mathcal{H}}}\left[ \frac{1}{D}\sum_{i=1}^{D}{\rm CE}\left(\mathbf{\hat{t}}^{(i)}, \mathbf{t}^{(i)}\right)\right], \label{eq:objective} \\
{\rm s.t.}~~& \mathbb{E}_{\boldsymbol{\mathcal{H}}}[P^{(i)}] \leq P_{\rm max}~~\forall i\!=\!1,\dots,D, \label{eq:power-constrint} \\
& \vartheta^l_n \in [0, 2\pi]~~\forall n\!=\!1,\dots,N,~\forall l\!=\!1,\dots,L, \label{eq:theta-constraint}
\end{align}
\end{subequations}
where $\mathbb{E}_{\boldsymbol{\mathcal{H}}}\left[\cdot\right]$ represents expectation over the channels, and ${\rm CE}(\mathbf{\hat{t}}^{(i)}, \mathbf{t}^{(i)}) \! \triangleq \! \sum_{c=1}^{C} [\mathbf{t}^{(i)}]_c \log([\mathbf{\hat{t}}^{(i)}]_c)$ assuming that $\mathbf{t}^{(i)}$ is the cross-entropy loss for multi-class classification, assuming that $\mathbf{t}^{(i)}$ is one-hot encoded and that $\mathbf{\hat{t}}^{(i)}$ can be interpreted as a vector of probabilities for each of the $C$ classes.

Once problem~\eqref{eq:problem} is solved, $f_{\mathbf{w}}(\cdot)$ may be used to infer the target values for previously unseen input data that follow the same distribution as those in $\mathcal{D}$.

\vspace{-0.3cm}
\section{Proposed Edge Inference Methodology}
\vspace{-0.1cm}
Since~\eqref{eq:problem}'s EI objective is an amortized cost function over the data set $\mathcal{D}$, we propose to make use of DL leveraging the high expressive power of the MINN framework~\cite{Stylianopoulos_GO}.
Our approach comprises the following four individual DNN modules:

\textit{The TX module}, $\mathbf{x}^{(i)}=f^{\rm t}_{\mathbf{w}_{\rm t}}(\mathbf{o}^{(i)})$, receives the observation and acts as a source encoder to provide the transmit signal $\mathbf{x}^{(i)}$ with normalized amplitude. It is noted that no explicit coding, modulation, or beamforming schemes are introduced, relying on the optimization of the parameter vector $\mathbf{w}_{\rm t}$ to implicitly learn advantageous strategies for the scenario at hand.

\textit{The power control module}, $P^{(i)} = f^{\rm p}_{\mathbf{w}_{\rm p}}(\mathbf{p}^{(i)})$, receives the current position of the RX, $\mathbf{p}^{(i)}=[x^{(i)},y^{(i)}]$, and outputs the power value to be used for $\mathbf{x}^{(i)}$'s transmission. Note that the input data are not used by this module allowing it to focus on beneficial power control strategies based solely on $\mathbf{p}^{(i)}$'s.

\textit{The channel module}, $\mathbf{y}^{(i)}=f^{\rm c}_{\boldsymbol{\theta}}(\mathbf{x}^{(i)}, P^{(i)}, \boldsymbol{\mathcal{H}}^{(i)})$, implements OTA the XL MIMO input-output relationship in~\eqref{eq:transmit-signal}, having as inputs the outputs of the latter TX and power modules. Recall that the channel is parametrized by the vector $\boldsymbol{\theta} \triangleq [\vartheta^l_n]^\top \! \in \! [0,2\pi]^{N_{\rm sim}}$ $\forall n,l$ containing all SIM elements' tunable responses, 
which are treated equivalently to trainable DNN parameters. To ensure that all phase shifts reside in the $[0,2\pi]$ range (satisfying~\eqref{eq:theta-constraint}), the operation $\boldsymbol{\theta} \! \gets \! \pi({\rm tan}^{-\!1}(\boldsymbol{\theta})\!+\!1)$ is applied to them.

\textit{The RX module}, $\mathbf{\hat{t}}^{(i)} = f^{\rm r}_{\mathbf{w}_{\rm r}}(\mathbf{y}^{(i)})$, decodes the received signals and outputs the target estimation. Similar to the TX module, no explicit decoding operations are imposed and no channel information is needed. Instead, the joint source-power-channel control of the previous modules is intended to provide an informative form for $\mathbf{\hat{y}}^{(i)}$ so that the final classification output can be correctly mapped.

By collecting the parameters of all latter modules into the vector $\mathbf{\bar{w}}\!\triangleq\![\mathbf{w}_{\rm t}^\top, \mathbf{w}_{\rm p}^\top, \boldsymbol{\theta}^\top,\mathbf{w}_{\rm r}^\top]^\top$, yields the proposed MINN:
\begin{equation}
    \mathbf{\hat{t}}^{(i)} = f^{\rm r}_{\mathbf{w}_{\rm r}}(f^{\rm c}_{\boldsymbol{\theta}}(f^{\rm t}_{\mathbf{w}_{\rm t}}(\mathbf{o}^{(i)}), f^{\rm p}_{\mathbf{w}_{\rm p}}(\mathbf{p}^{(i)}), \boldsymbol{\mathcal{H}}^{(i)})),
\end{equation}
which is trained to solve a relaxed version of~\eqref{eq:problem}. In particular, we incorporate constraint~\eqref{eq:power-constrint} into \eqref{eq:objective}'s objective yielding the following design problem formulation for $\mathbf{\bar{w}}$:
\begin{equation}\label{eq:relaxed-objective}
    \min_{\mathbf{\bar{w}}}\mathcal{\hat{L}}(\mathbf{\bar{w}}) \triangleq \mathbb{E}_{\boldsymbol{\mathcal{H}}} \left[ \sum_{i=1}^{D}\!{\rm CE}\left(\mathbf{\hat{t}}^{(i)}, \mathbf{t}^{(i)}\right) + \gamma P^{(i)} \right], 
\end{equation}
where $\gamma$ denotes the Lagrange penalty that is empirically determined to satisfy the $P_{\rm max}$ constraint. In the remainder, we make the assumption that $(\mathbf{o}^{(i)},\mathbf{t}^{(i)})$ and $\boldsymbol{\mathcal{H}}^{(i)}$ are statistically independent, requiring the data to have no relation to the wireless environment.
This assumption ensures that random pairs of $(\mathbf{o}^{(i)},\mathbf{t}^{(i)})$ and $\boldsymbol{\mathcal{H}}^{(i)}$, sampled independently, provide unbiased estimators for the expectations of $\mathcal{\hat{L}}(\mathbf{\bar{w}})$~\cite{Robbins_Monro}.
Therefore, backpropagation through stochastic gradient descent~\cite{Rumelhart86} is used on~\eqref{eq:relaxed-objective} to find approximately optimal $\mathbf{\bar{w}}$ values.
Due to lack of space, we skip the derivation of the partial derivatives for the weights of each of the DNN modules using the chain rule, since all operations involve analytically differentiable functions that can be computed through automatic differentiation software~\cite{PyTorch}.
Related derivations w.r.t. $\boldsymbol{\theta}$ have been presented in~\cite{Stylianopoulos_GO}.
Note that none of the digital DNN modules require knowledge of the channel matrices during their forward pass (and therefore during deployment) since $f^{\rm c}_{\boldsymbol{\theta}}(\cdot)$ is implemented OTA.
Nevertheless, an analytical model for $f^{\rm c}_{\boldsymbol{\theta}}(\cdot)$ and knowledge of $\boldsymbol{\mathcal{H}}^{(i)}$'s are still needed during training to backpropagate the gradients to the TX and power modules.

\vspace{-0.2cm}
\section{Numerical Results}
\vspace{-0.2cm}
For the numerical evaluation, we have used a set up of static clutter where channel variations rise from sampling different user positions within the area of interest.
The TX was located at $\mathbf{p}_{\rm t}\!=\![0m,0m,4m]$, the first element of the SIM at $\mathbf{p}_{\rm m}\!=\![0.5m,0m,4m]$, while the user was randomly sampled in the bounding box of $[1m,-5m,0.5m]\!\times \![25m,5m,2.5m]$.
All antenna elements were placed at $\lambda/2$ distances, with $\lambda\!=\!10^{-2}~{ }$, $d_M\!=\!10\lambda$, $S_M\!=\!\lambda^2/4$, and $\sigma^2\!=\!-90~{\rm dBm}$.
Unless otherwise specified, the values
$N_{\rm t}\!=\!16$, $N_{\rm r}\!=\!8$, $N\!=\!10\!\times\!10$ (planar MSs), $L\!=\!4$, $K\!=\!20$, and $\gamma\!=\!10^{-2}$ were used.

We have performed experiments on MNIST classification~\cite{MNIST}.
A three-layer Convolutional Neural Network (CNN) followed by three linear layers has been trained to $0.991$ classification accuracy, which presents an upper bound.
A three-layer CNN of $32$ channels followed by two linear layers of $128$ units constituted $f^{\rm t}_{\mathbf{w}{\rm t}}(\cdot)$, while $f^r_{\mathbf{w}{\rm r}}(\cdot)$ used three linear layers of $128$ units.
Three layers of $16$ units are used for $f^{\rm p}_{\mathbf{w}{\rm p}}(\cdot)$.
Rectified Linear Units (ReLUs) were used as activations everywhere apart from the final layers.
Each training procedure lasted for $150$ epochs using the Adam optimizer~\cite{Adam} with learning rate of $10^{-3}$, upon sampling $10^4$ training and $10^3$ testing channel realizations.
To speed up convergence, it was found beneficial to start training the power module after the $50$-th epoch setting $P^{(i)}\!=\!20~{\rm dBm}$ at the initial epochs.
The maximum test accuracies alongside the mean power used by the network during the corresponding epoch are reported in the following figures.
Results are averaged out over $5$ different random seeds.

\begin{figure}[t]
    \centering
    \includegraphics[width=1.0\linewidth]{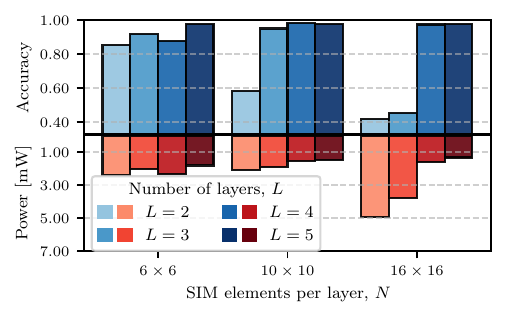}
    \vspace{-0.8cm}
    \caption{Achieved EI accuracy and corresponding TX power of the proposed MINN model incorporating power control under different numbers of SIM layers and unit elements per layer.}
    \label{fig:acc_pow_vs_N_L}
\end{figure}

\begin{figure}[t]
    \centering
    \includegraphics[width=1.0\linewidth]{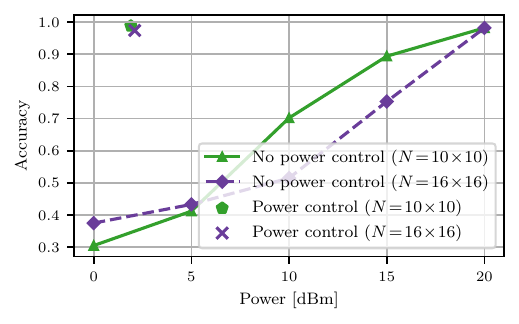}
    \vspace{-0.8cm}
    \caption{Achieved EI accuracy of different MINN versions with no power control modules over predetermined TX power levels, and comparison to achieved accuracy-power performance with MINNs realizing power control.}
    \label{fig:acc_vs_pow_no_power_control}
\end{figure}

The first part of the investigations concerns the required numbers of SIM layers and MS unit elements for successful EI, as shown in~\autoref{fig:acc_pow_vs_N_L}. As observed, the EI problem can be effectively solved even by deeper SIM of $36$ units per layer, achieving $0.988$ classification accuracy, which approaches that of the fully digital DNN benchmark.
Larger MSs (i.e., SIM with wider layers) require extensive training periods, often result in suboptimal performance, and provide limited gains in terms of power consumption.
The number of SIM layers appears to be a more important factor, reaffirming a consensus in the DL community that deeper DNNs are more effective than shallow ones with large width~\cite{ShallowDeepNNs}.
To particularly assess the importance of the power control module, we have implemented baseline MINNs with the same structure, but replacing the power module with predefined TX power levels, as shown in~\autoref{fig:acc_vs_pow_no_power_control}. It is shown that the power required to approach the previous MINN performance in the absence of power modules reaches $20~{\rm dBm}$.
It can thus be inferred that power control is a crucial component for achieving energy efficiency in MINNs, offering approximately $18~{\rm dB}$ improvement in the particular case.

\begin{figure}[t]
    \centering
    \includegraphics[width=1.0\linewidth]{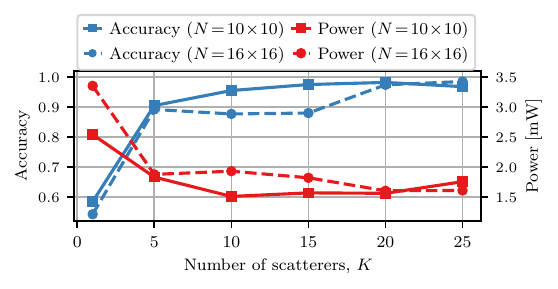}
    \vspace{-0.8cm}
    \caption{Achieved EI accuracy and power control of MINN versions over different channel richness levels.}
    \label{fig:acc_pow_vs_scatterers}
\end{figure}

\begin{figure}[t]
    \centering
    \subfloat[\vspace{-0.2cm}EI Accuracy\label{fig:heatmap_acc}]{

        \includegraphics[width=0.48\linewidth]{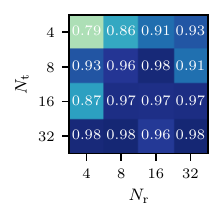}
    }~%
    \subfloat[\vspace{-0.2cm}Power \lbrack mW\rbrack \label{fig:heatmap_pow}]{
        \includegraphics[width=0.48\linewidth]{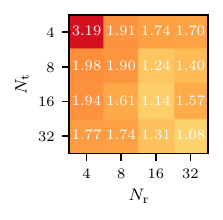}
    }
    \caption{Performance with the proposed MINN over different numbers of TX $(N_{\rm t})$ and RX $(N_{\rm r})$ antennas  ($N=10\times10$).}
    \label{fig:heatmap_Nt_Nr}
    
\end{figure}

\begin{figure}[t]
    \centering
    \includegraphics[width=1.0\linewidth]{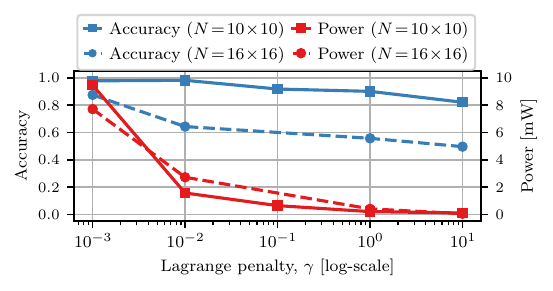}
    \vspace{-0.8cm}
    \caption{Performance with the proposed MINN over different values of the Lagrange parameter controlling the accuracy-power consumption trade-off.}
    \label{fig:acc_pow_vs_weight}
\end{figure}

We next investigate the effects of scattering in the environment, as shown in~\autoref{fig:acc_pow_vs_scatterers}.
The channel responses have been normalized over the number of scatterers $K$ in~\eqref{eq:channel-tx-rx} and~\eqref{eq:channel-ris-rx} to avoid spurious power increases under rich scattering. As observed, few reflecting paths do not provide enough diversity in the channel matrices to be effectively controlled for OTA DL computations, leading to lower EI accuracy and higher TX power. For $K\! \geq \!15$, the performance metrics converge to their optimal values.  
An investigation over different MIMO configurations is given in~\autoref{fig:heatmap_Nt_Nr}.
As expected, at larger antenna systems both performance indicators improve, but overall, the MINNs are robust to various antenna numbers.
Power consumption is more strongly affected, while increasing the number of TX antennas from small values leads to greater improvements.
Finally, accuracy-power trade-offs over a wide range of the Lagrange penalty parameter are presented in~\autoref{fig:acc_pow_vs_weight}. It can be seen that smaller values result in high power budgets up to $10~{\rm mW}$ with only marginal improvements in accuracy, while larger values help minimize the transmission power; however, they incur an accuracy decrease of up to $30\%$.
This behavior illustrates that the proposed methodology provides a direct way of configuring the MINN to balance the constraints of intended applications.

\vspace{-0.1cm}
\section{Conclusion}
\vspace{-0.2cm}
In this paper, SIMs were integrated as part of a TX-channel-RX DNN with a power control module that performs EI. Instead of being treated as a source of distortion to be negated, the SIM-parameterized channel was controlled for OTA DL.
The proposed E2E model is differentiable and trainable under dynamic fading conditions. Once trained, the SIM responses remain fixed, which has the potential to simplify their manufacturing. Our results showcased that the proposed methodology  is robust across various parameters providing direct control of the accuracy-power trade-off. It was also demonstrated that the power control module achieves up to $18~{\rm dB}$ decrease in TX power while maintaining the same accuracy levels.

\FloatBarrier
\vfill\pagebreak
\newpage
\bibliographystyle{IEEEbib}
\bibliography{refs}

\end{document}